\documentclass[aip,jap,unsortedaddress,superscriptaddress,reprint]{revtex4-1}
\usepackage{hyperref}
\usepackage{graphicx}
\usepackage{amsmath}
\usepackage{amssymb}
\usepackage{times,mathptmx}
\usepackage{dcolumn}
\usepackage{ifthen}
\usepackage[dvipsnames,usenames]{pstricks}
\usepackage{color}
\usepackage{ulem}             
\usepackage{proof}

\newcommand{\beginsupplement}{%
        \setcounter{table}{0}
        \renewcommand{\thetable}{S\arabic{table}}%
        \setcounter{equation}{0}
        \renewcommand{\theequation}{S\arabic{equation}}%
        \setcounter{figure}{0}
        \renewcommand{\thefigure}{S\arabic{figure}}%
     }

\definecolor{rot}{rgb}{1,0,0}

\newif\ifdel
\deltrue               %

\begin{document}

\title{Pure spin current transport in gallium doped zinc oxide}
\author{Matthias Althammer}
\email{matthias.althammer@wmi.badw.de}
\affiliation{Walther-Mei{\ss}ner-Institut, Bayerische Akademie der Wissenschaften, 85748 Garching, Germany}
\affiliation{Physik-Department, Technische Universit\"at M\"unchen, 85748 Garching, Germany}
\author{Joynarayan Mukherjee}
\affiliation{Department of Physics, Nano Functional Materials Technology Centre and Materials Science Research Centre, Indian Institute of Technology Madras, Chennai, Tamil Nadu 600036, India}
\author{Stephan Gepr\"ags}
\affiliation{Walther-Mei{\ss}ner-Institut, Bayerische Akademie der Wissenschaften, 85748 Garching, Germany}
\author{Sebastian T.~B.~Goennenwein}
\affiliation{Walther-Mei{\ss}ner-Institut, Bayerische Akademie der Wissenschaften, 85748 Garching, Germany}
\author{Matthias Opel}
\affiliation{Walther-Mei{\ss}ner-Institut, Bayerische Akademie der Wissenschaften, 85748 Garching, Germany}
\author{M.S. Ramachandra Rao}
\affiliation{Department of Physics, Nano Functional Materials Technology Centre and Materials Science Research Centre, Indian Institute of Technology Madras, Chennai, Tamil Nadu 600036, India}
\author{Rudolf Gross}
\affiliation{Walther-Mei{\ss}ner-Institut, Bayerische Akademie der Wissenschaften, 85748 Garching, Germany}
\affiliation{Physik-Department, Technische Universit\"at M\"unchen, 85748 Garching, Germany}
\affiliation{Nanosystems Initiative Munich (NIM), 80799 M\"unchen, Germany}

\date{\today}
\begin{abstract}
We study the flow of a pure spin current through zinc oxide by measuring the spin Hall magnetoresistance (SMR) in thin film trilayer samples consisting of bismuth-substituted yttrium iron garnet (Bi:YIG), gallium-doped zinc oxide (Ga:ZnO), and platinum. We investigate the dependence of the SMR magnitude on the thickness of the Ga:ZnO interlayer and compare to a Bi:YIG/Pt bilayer. We find that the SMR magnitude is reduced by almost one order of magnitude upon inserting a Ga:ZnO interlayer, and continuously decreases with increasing interlayer thickness. Nevertheless, the SMR stays finite even for a $12\;\mathrm{nm}$ thick Ga:ZnO interlayer. These results show that a pure spin current indeed can propagate through a several nm-thick degenerately doped zinc oxide layer. We also observe differences in both the temperature and the field dependence of the SMR when comparing tri- and bilayers. Finally, we compare our data to predictions of a model based on spin diffusion. This shows that interface resistances play a crucial role for the SMR magnitude in these trilayer structures.
\end{abstract}
\maketitle

\begin{figure*}[tb]
 \includegraphics[width=170mm]{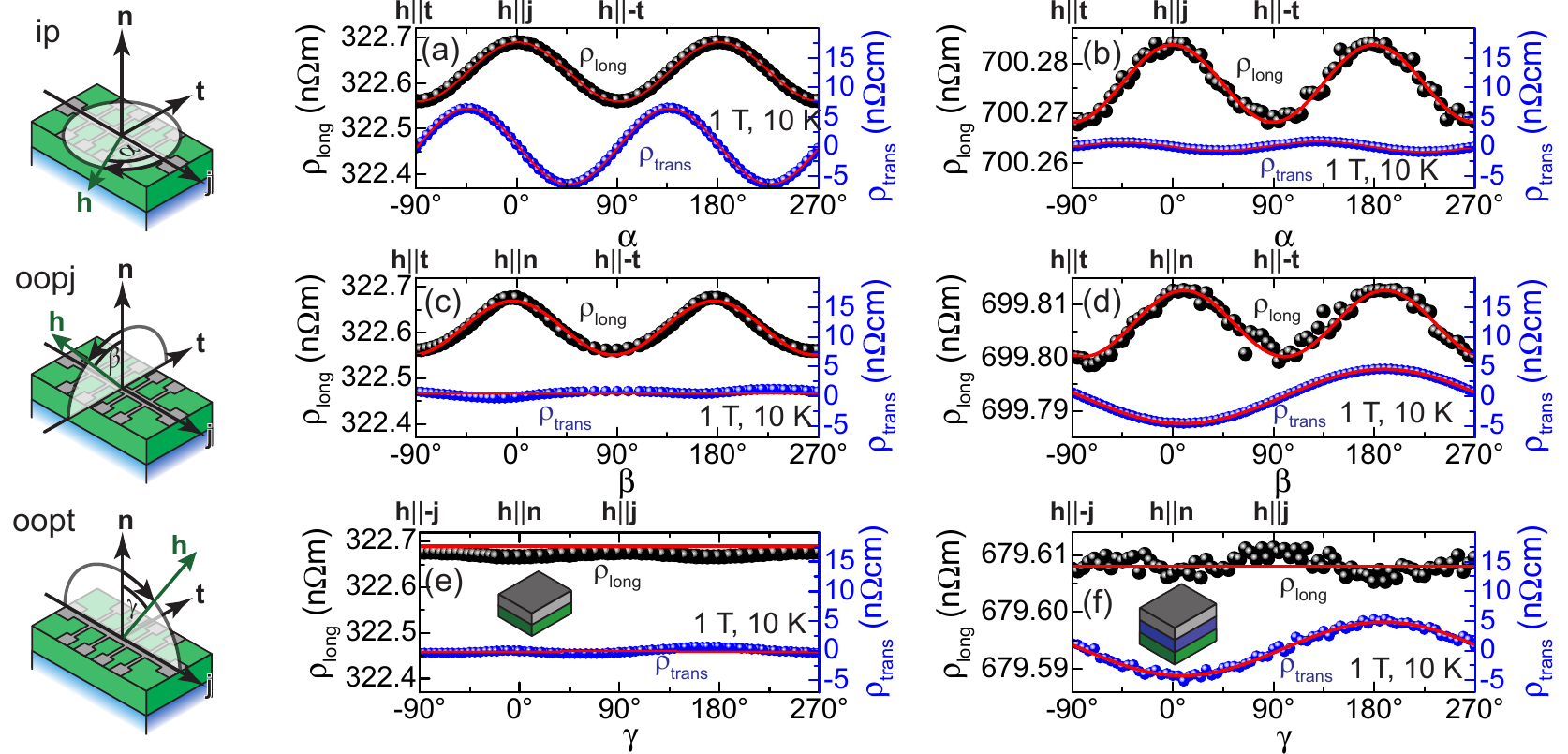}\\
 \caption[ADMR results for Bilayer and Trilayer]{ADMR data of a Bi:YIG(54)/Pt(4) bilayer (panels (a), (c), and (e)) and a Bi:YIG(54)/Ga:ZnO(8)/Pt(4) trilayer (panels (b), (d), and (f)) sample grown on YAG (111) substrates. The data has been recorded at $10\;\mathrm{K}$. The three orthogonal rotation planes for the magnetic field are sketched in the left column. In the plot, black and blue symbols represent the experimental data of the longitudinal $\rho_\mathrm{long}$ and transverse resistivity $\rho_\mathrm{trans}$, respectively. The red lines are simulations using Eqs.~\eqref{Eq:SMRRhoLong},\eqref{Eq:SMRRhoTrans}.}
  \label{figure:ADMR_BiYIG}
\end{figure*}

The generation and detection of pure spin currents, i.e.~of net flows of angular (spin) momentum without accompanying charge currents, has been intensively studied in theory and experiments. Prominent spin current based phenomena are spin pumping~\cite{Ando2010,Czeschka2011}, the spin Seebeck effect~\cite{Uchida2008,Uchida2010,Weiler2012} and the spin Hall magnetoresistance~\cite{nakayama_spin_2013,Hahn2013,chen_theory_2013,althammer_quantitative_2013,Vlietstra2013} (SMR). The SMR manifests itself in ferromagnetic insulator (FMI) / normal metal (NM) bilayer samples, as a dependence of the electric resistance of the NM on the orientation of the magnetization in the FMI. The magnitude of the SMR depends on the size of the spin Hall angle in the NM. In a FMI/NM bilayer system, the longitudinal ($\rho_\mathrm{long}$) and transverse resistivity ($\rho_\mathrm{trans}$) of the NM can be written as~\cite{nakayama_spin_2013,chen_theory_2013,althammer_quantitative_2013,Ganzhorn2016}
\begin{eqnarray}
\rho_\mathrm{long}&=\rho_0+\rho_1 (m_\mathrm{t})^2,
\label{Eq:SMRRhoLong}\\
\rho_\mathrm{trans}&=\rho_2 m_\mathrm{n}-\rho_1 (m_\mathrm{j} m_\mathrm{t})^2,
\label{Eq:SMRRhoTrans}
\end{eqnarray}
where $m_\mathrm{j}$,$m_\mathrm{t}$, and $m_\mathrm{n}$ are the projections of the net magnetization unit vector $\mathbf{m}=\mathbf{M}/M$ determined by the magnetization orientation in the FMI on the directions $\mathbf{j}$ and $\mathbf{t}$ parallel and perpendicular to the current direction, respectively, and the film normal $\mathbf{n}$ (cf.~Fig.~\ref{figure:ADMR_BiYIG}). The resistivity parameters $\rho_i$ depend on the material parameters of the hybrid structure. A detailed analysis allows to extract the spin Hall angle $\theta_\mathrm{SH,NM}$ and the spin diffusion length $\lambda_\mathrm{sf,NM}$ of the NM, if the spin mixing conductance $g_{\uparrow\downarrow}$ at the FMI/NM interface is known.~\cite{althammer_quantitative_2013,weiler_experimental_2013}

Up to now, most SMR studies were based on FMI/NM bilayer structures. Only in a few experiments an additional conductive interlayer was inserted at the FMI/NM interface to rule out the contribution of a proximity-polarized NM layer.~\cite{Geprags2012,althammer_quantitative_2013,weiler_experimental_2013,Miao2014} However, trilayer structures also allow to study the transport of pure spin currents in interlayer materials with a negligible spin Hall angle. In particular, the effect of interlayer resistivity and spin diffusion length onto the SMR has not yet been investigated. Moreover, a quantitative comparison between a spin diffusion theory model and experiment is still missing for such trilayer systems. In this letter, we present SMR experiments conducted on bismuth-substituted yttrium iron garnet (Bi$_{0.3}$Y$_{2.7}$Fe$_5$O$_{12}$, Bi:YIG) / platinum (Pt) bilayers and Bi:YIG / gallium doped zinc oxide (Ga$_{0.01}$Zn$_{0.99}$O, Ga:ZnO) / Pt trilayers. Our results demonstrate pure spin current transport across a degenerately doped, several nm-thick ZnO interlayer, resulting in a sizeable SMR effect. In addition, we exploit the tunability of the resistivity and spin diffusion length in Ga:ZnO with temperature to quantitatively compare our experimental data with the predictions of a model based on spin diffusion. This comparison suggests that interface resistances and their possible spin selectivity may play a crucial role for the SMR in these trilayer structures.

The samples have been grown \textit{in-situ}, without breaking the vacuum, on (111)-oriented yttrium aluminium garnet (YAG) substrates in an ultra high vacuum deposition system. Bi:YIG and Ga:ZnO layers were deposited via laser-molecular beam epitaxy (laser-MBE) from stoichiometric targets in an oxygen atmosphere. For Bi:YIG we applied an energy density at the target of $ED_\mathrm{L}=1.5\;\mathrm{J/cm^2}$, an oxygen partial pressure of $p_\mathrm{O_2}=25\;\mathrm{\mu bar}$, and a substrate temperature of $T_\mathrm{sub}=450\;\mathrm{^\circ C}$ (see Ref.~\onlinecite{althammer_quantitative_2013} for more details). For Ga:ZnO we used: $ED_\mathrm{L}=1\;\mathrm{J/cm^2}$, $p_\mathrm{O_2}=1\;\mathrm{\mu bar}$, $T_\mathrm{sub}=400\;\mathrm{^\circ C}$ (See Ref.~\onlinecite{Althammer2012} for more details). Pt was deposited via electron-beam evaporation at a base pressure of $1\times10^{-8}\;\mathrm{mbar}$ at room temperature. The \textit{in-situ} deposition process ensures very clean interfaces.

From the structural and magnetic characterization of a Bi:YIG(54)/Pt(4) bilayer and a Bi:YIG(54)/Ga:ZnO(8)/Pt(4) trilayer, where the numbers in parentheses give the  film thicknesses in nm, we conclude that the structural and the magnetic properties of the bi- and trilayer sample are nearly identical (see supplemental materials). This suggests that the influence of the Ga:ZnO deposition on the magnetic properties of the Bi:YIG layer is negligible.

For electrical transport measurements, the samples have been patterned into Hall bar-shaped mesastructures ($80\;\mathrm{\mu m}$ wide, $800\;\mathrm{\mu m}$ long) via photolithography and Ar-ion milling, and then mounted in a superconducting magnet ($\mu_0H \leq 7 \;\mathrm{T}$) cryostat ($5\;\mathrm{K}\leq T\leq 300\;\mathrm{K}$). Resistivity data have been taken using a DC current reversal technique. From the measured longitudinal and transverse voltage we then calculated $\rho_\mathrm{long}$ and $\rho_\mathrm{trans}$.~\cite{althammer_quantitative_2013,Schreier2013} We performed angle-dependent magnetoresistance (ADMR) measurements, where the direction of the applied magnetic field of constant magnitude is rotated within three orthogonal planes as illustrated in the left column of Fig.~\ref{figure:ADMR_BiYIG}: in the film plane (ip), in the plane perpendicular to the $\mathbf{j}$-direction (oopj), and in the plane perpendicular to the $\mathbf{t}$-direction (oopt). Although the smallest field magnitude is below the full saturation field of Bi:YIG of $2\;\mathrm{T}$, it is sufficient to investigate the main angular dependence.

We first discuss the results obtained from ADMR experiments at $T=10\;\mathrm{K}$ and $\mu_0 H=1\;\mathrm{T}$ for our bilayer and trilayer sample, as shown in Fig.~\ref{figure:ADMR_BiYIG}. For the bilayer we measure the typical fingerprint expected for SMR~\cite{althammer_quantitative_2013}: For the ip rotation plane we observe a $\sin^2(\alpha)$-dependence of $\rho_\mathrm{long}$ and a $\sin(\alpha)\cos(\alpha)$-dependence of $\rho_\mathrm{trans}$ (see Fig.~\ref{figure:ADMR_BiYIG}(a)), for the oopj rotation plane we observe a $\sin^2(\beta)$-dependence of $\rho_\mathrm{long}$ (see Fig.~\ref{figure:ADMR_BiYIG}(c)), while we do not observe any sizeable angular-dependence of $\rho_\mathrm{long}$ in the oopt rotation plane (see Fig.~\ref{figure:ADMR_BiYIG}(e)). Furthermore, $\rho_\mathrm{trans}$ only shows a very small cosine dependence in the oopj and oopt rotation planes, due to the nearly vanishing ordinary Hall coefficient (OHC) of Pt thin films at low temperatures.~\cite{Meyer2015} We note that the remaining $\sin^2$-dependencies, for $\rho_\mathrm{long}$ in oopt and for $\rho_\mathrm{trans}$ in oopj and oopt configuration, can be explained by a non-vanishing SMR contribution in these rotation planes due to a small tilting ($\leq 3^\circ$) of the actual rotation plane with respect to the surface normal because of experimental limitations in the sample mounting. For the trilayer sample the angular dependence looks qualitatively the same (Fig.~\ref{figure:ADMR_BiYIG}(b,d,f)).
As the observed ADMR data reflects the symmetry expected for SMR, we can safely assume the SMR as the only cause for the magnetoresistance in the bilayer as well as for the trilayer.

\begin{figure}[b]
 \includegraphics[width=85mm]{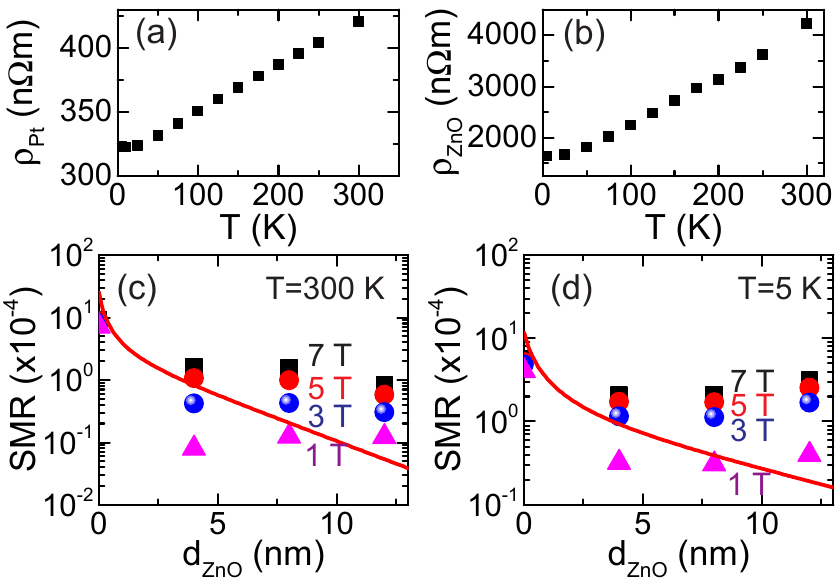}\\
 \caption[thickness dependence SMR for trilayer samples]{Extracted temperature dependence of the resistivity of Pt (a) from the bilayer and the resistivity of Ga:ZnO (b) extracted from the trilayer ADMR data by using the resistivity of Pt extracted from the bilayer. For Pt and ZnO the resistivity increases with increasing temperature. SMR amplitude as a function of Ga:ZnO thickness at $300\;\mathrm{K}$ (c) and $5\;\mathrm{K}$(d). The red line is a spin diffusion based simulation of the SMR amplitude.}
  \label{figure:SMR_thick_BiYIG}
\end{figure}

Quantitatively, however, there are differences. In the trilayer, $\rho_\mathrm{long}$ is about 2 times larger than in the bilayer, which can be explained by the one order of magnitude larger resistivity of the ZnO layer as compared to the bare Pt layer and the effective average of resistivity from Pt and Ga:ZnO for the trilayer (see Fig.~\ref{figure:SMR_thick_BiYIG}(b) for the extracted resisitivity of the ZnO layer). For further quantitative comparison, we simulated the SMR response for $\rho_\mathrm{long}$ and $\rho_\mathrm{trans}$ (red lines in the graphs) by using Eqs.~\eqref{Eq:SMRRhoLong} and \eqref{Eq:SMRRhoTrans}, while assuming that the magnetization orientation is always parallel to the external magnetic field and including the ordinary Hall effect as a contribution parameterized by a field dependence of $\rho_2$. From the simulation we extract the SMR amplitude $\mathrm{SMR}=|\rho_1/\rho_0|$. For the bilayer we obtain $\mathrm{SMR}=4.0\times10^{-4}$, which agrees nicely with our previous results in YIG/Pt hybrids.~\cite{althammer_quantitative_2013,Meyer2014} For the trilayer we find $\mathrm{SMR}=2.2\times10^{-5}$, which is about an order of magnitude smaller. There are two obvious reasons for the decrease in SMR amplitude upon the insertion of a Ga:ZnO interlayer. First, in the trilayer the Ga:ZnO layer acts as a resistive shunt for the Pt layer thereby reducing the SMR amplitude. Second, part of the spin current generated in the Pt layer is lost when diffusing across the Ga:ZnO layer. In other words, the Ga:ZnO only acts as a parallel and spin Hall-inactive resistor for the SMR. Moreover, when comparing the transverse resistivity for both samples, the amplitude of the ADMR signal is different in all rotation planes for the two different samples (please note that the same scale has been used for $\rho_\mathrm{trans}$ in all graphs of Fig.~\ref{figure:ADMR_BiYIG}). For the ip rotation plane the bilayer has a much larger amplitude than for the trilayer. This is expected due to the larger SMR in the bilayer, which is the only contribution to the ADMR for this rotation plane. For the oopj and oopt rotation planes, however, the $\rho_\mathrm{trans}$ amplitude is larger for the trilayer. This can be traced back to the contribution of the ordinary Hall effect to the ADMR data: for the bilayer the OHC is rather small~\cite{Meyer2015}($\approx0.1\;\mathrm{m\Omega/T}$), while for the trilayer it is an effective average of the Pt layer and the Ga:ZnO interlayer, resulting in a larger OHC. From the field dependence of $\rho_2$ for the trilayer, we extract an OHC of $30\;\mathrm{m\Omega/T}$ for the Ga:ZnO, corresponding to a carrier concentration of $2\times10^{21}\mathrm{cm^{-3}}$. This value nicely agrees with control measurements on blanket Ga:ZnO layers (see supplemental material).

\begin{figure}[tb]
 \includegraphics[width=85mm]{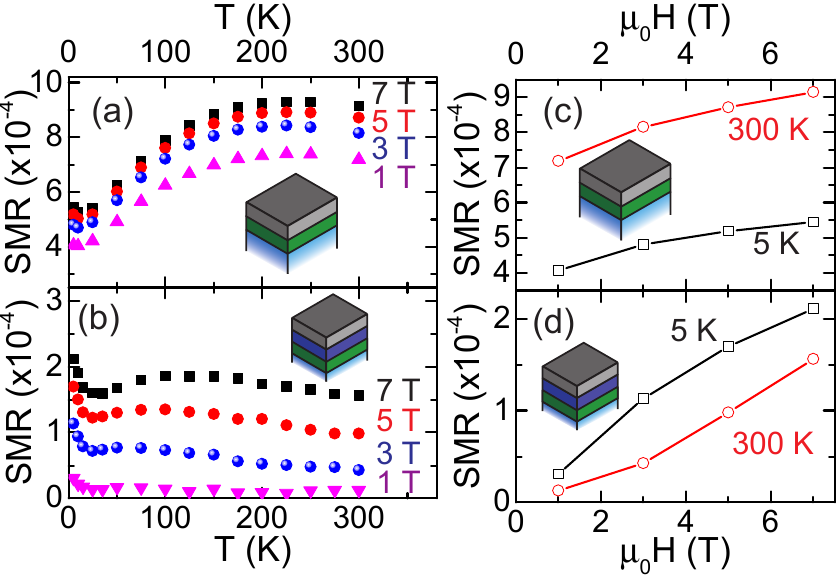}\\
 \caption[Temperature dependence of SMR]{Temperature dependence of the $\mathrm{SMR}$ signal extracted from ADMR measurements for (a) a Bi:YIG(54)/Pt(4) bilayer and (b) a Bi:YIG(54)/Ga:ZnO(8)/Pt(4) trilayer at magnetic fields of $1, 3, 5, 7\;\mathrm{T}$. Magnetic field dependence of $\mathrm{SMR}$ from ADMR measurements for the bilayer (c) and the trilayer (d) at $5\;\mathrm{K}$ (black squares) and $300\;\mathrm{K}$ (red circles).}
  \label{figure:SMR_Temp_BiYIG}
\end{figure}

To get a deeper insight into the SMR of the trilayer we measured the temperature and field dependence and compare it to the results for the bilayer sample. To this end, we conducted ADMR measurements in the ip rotation plane for $5\leq T \leq 300\;\mathrm{K}$ and $\mu_0 H=1,3,5,$ and $7\;\mathrm{T}$ and extracted the $\mathrm{SMR}$ by fitting the data using Eq.\eqref{Eq:SMRRhoLong}. The result is shown in Fig.~\ref{figure:SMR_Temp_BiYIG}.
We first discuss the temperature dependence for the bilayer sample in Fig.~\ref{figure:SMR_Temp_BiYIG}(a). The $\mathrm{SMR}$ shows a maximum around $T=225\;\mathrm{K}$ and then decreases with decreasing temperature. This observation nicely agrees with results on YIG/Pt hybrids~\cite{Meyer2014,Aqeel2015} and can be attributed to the temperature dependence of $\theta_\mathrm{SH,Pt}$.~\cite{Meyer2014} However, the temperature dependence is clearly different for the trilayer sample (Fig.~\ref{figure:SMR_Temp_BiYIG}(b)). Here, the $\mathrm{SMR}$ is only weakly changing with temperature, displays an upturn towards low temperatures and reaches its maximum value for $5\;\mathrm{K}$. We may explain this upturn by two contributions. On the one hand, the spin diffusion length for Ga:ZnO increases with decreasing temperature as it is dominated by the D'yakonov-Perel' mechanism for spin dephasing.~\cite{Ghosh2005,Althammer2012} This leads to an increase of the $\mathrm{SMR}$ and a saturation at low temperatures. On the other hand, due to the laser-MBE deposition of Ga:ZnO an intermixing of Bi:YIG and Ga:ZnO at the interface can occur, which may lead to the formation of isolated paramagnetic moments at the interface between Ga:ZnO and Bi:YIG. These isolated paramagnetic moments align parallel to the external magnetic field at low temperatures, which would also result in an increase of $\mathrm{SMR}$ with decreasing temperatures. The temperature dependent data of the bilayer also shows a weak plateau at low temperatures, which could be related to isolated paramagnetic moments. However, more detailed investigations of the interface properties will be necessary in the future in order to really unravel the underlying physics. 

The evolution of the magnetic field dependence of the $\mathrm{SMR}$ with temperature for the bilayer and the trilayer sample is also different. For the bilayer the relative increase of the $\mathrm{SMR}$ from $1\;\mathrm{T}$ to $7\;\mathrm{T}$ is $37\%$ at $5\;\mathrm{K}$ and $27\%$ at $300\;\mathrm{K}$(Fig.~\ref{figure:SMR_Temp_BiYIG}(c)). For the trilayer we obtain $600\%$ at $5\;\mathrm{K}$ and $1500\%$ at $300\;\mathrm{K}$(Fig.~\ref{figure:SMR_Temp_BiYIG}(d)). The much stronger field dependence observed for the trilayer sample again might be explained by the presence of isolated paramagnetic moments at the Bi:YIG/Ga:ZnO interface. 
From our bilayer data it is not directly possible to pin point the origin for the observed field dependence. 
Hanle magnetoresistance as proposed by V{\'{e}}lez \textit{et al.}~\cite{Vlez2016} can be ruled out, as this should yield a quadratic field dependence of the $\mathrm{SMR}$, while we observe a linear/square root dependence in our data. Clearly, more detailed studies are necessary to explain the observed field dependence.

To analyze the pure spin current transport in Ga:ZnO we first extract the resistivity of Pt ($\rho_\mathrm{Pt}(T)$) from the $\rho_\mathrm{long}(T)$ of the bilayer and then determine the resistivity of Ga:ZnO ($\rho_\mathrm{ZnO}(T)$) from the measured $\rho_\mathrm{long}(T)$ of the trilayer by assuming a parallel conductance model and the same $\rho_\mathrm{Pt}(T)$ as for the bilayer (see Fig.~\ref{figure:SMR_thick_BiYIG}(a) and (b)). For both materials we observe a decrease in resistivity with decreasing temperature, while $\rho_\mathrm{ZnO}$ is about an order of magnitude larger than $\rho_\mathrm{Pt}$. From this we conclude that the Ga:ZnO layer just behaves like a dirty metal due to degenerate doping.

In a next step we model the dependence of the $\mathrm{SMR}$ on the Ga:ZnO thickness ($d_\mathrm{ZnO}$) using the SMR theory approach~\cite{chen_theory_2013} and adding the spin diffusion through the Ga:ZnO to the theory model, while neglecting any interface resistance effects, i.e. assuming continuity of the spin-dependent electrochemical potential at the Pt/Ga:ZnO interface (see supplemental materials). We note that ADMR measurements on bare Bi:YIG/Ga:ZnO reference samples yield no SMR response within our experimental resolution (for a $8\;\mathrm{nm}$ thick Ga:ZnO layer on Bi:YIG, we find $\mathrm{SMR}\leq 8\times 10^{-6}$ for $5\;\mathrm{K}\leq T \leq 300\;\mathrm{K}$ and $\mu_0 H\leq 7\;\mathrm{T}$), such that we assume $\theta_\mathrm{SH,ZnO}= 0$. From this model, two parameters define the $d_\mathrm{ZnO}$ dependence of the $\mathrm{SMR}$: $\rho_\mathrm{ZnO}$ and the spin diffusion length $\lambda_\mathrm{sf,ZnO}$ in the Ga:ZnO layer. The resistivity $\rho_\mathrm{ZnO}$ has a twofold implication. On the one hand, it determines the amount of charge current flowing through Pt and, hence, the amount of spin current generated via the spin Hall effect in Pt. On the other hand, it parameterizes the amount of spin current diffusing through Ga:ZnO due to the gradient in the spin-dependent electrochemical potential. To test whether or not this simple model explains our experimental findings, we simulated $\mathrm{SMR}$ at $5\;\mathrm{K}$ and $300\;\mathrm{K}$ and compared it to our experimental data as shown in Fig.~\ref{figure:SMR_thick_BiYIG}(c) and (d). We also included ADMR data obtained for Bi:YIG(54)/Ga:ZnO(12)/Pt(9) and Bi:YIG(54)/Ga:ZnO(4)/Pt(7) trilayers, which showed a similar surface roughness determined from x-ray reflectometry and thus similar interface properties. For better comparison, we renormalized the resistivity data to a Pt thickness of $4\;\mathrm{nm}$ using our previous results~\cite{Meyer2014}. For the simulation we used $\rho_\mathrm{Pt}$ and $\rho_\mathrm{ZnO}$ from Fig.~\ref{figure:SMR_thick_BiYIG}(a) and (b), while we chose temperature-independent values of $g_{\uparrow\downarrow}=1\times10^{19}\;\mathrm{m^{-2}}$ and $\lambda_\mathrm{sf,Pt}=1.5\;\mathrm{nm}$ from Refs.~\onlinecite{althammer_quantitative_2013,weiler_experimental_2013,Meyer2014}. For $\theta_\mathrm{SH,Pt}$ we used $0.11$ at $300\;\mathrm{K}$ and $0.07$ at $5\;\mathrm{K}$ from Ref.~\onlinecite{Meyer2014} as well as $4\;\mathrm{nm}$ at $300\;\mathrm{K}$ and $12\;\mathrm{nm}$ at $5\;\mathrm{K}$ for $\lambda_\mathrm{sf,ZnO}$ from Ref.~\onlinecite{Althammer2012}. Our very simple model can only reproduce the general trend of the measurements (see Fig.~\ref{figure:SMR_thick_BiYIG}(c) and (d)). The differences between simulation and data evident from Fig.~\ref{figure:SMR_thick_BiYIG} most likely originate from the crucial role of a spin dependent interface resistance~\cite{Althammer2012} at the Pt/Ga:ZnO interface and a possible change of $g_{\uparrow\downarrow}$ when going from a Bi:YIG/Pt to a Bi:YIG/Ga:ZnO interface. However, to extract these parameters from our measurements, a more systematic study is necessary, which goes beyond the scope of this paper.

In summary, we experimentally investigated the flow of a pure spin current through a Ga:ZnO layer by utilizing the SMR in Bi:YIG/Ga:ZnO/Pt trilayer thin film samples. Our results show the possibility to transfer a pure spin current through a degenerately doped ZnO interlayer. 
Our results also highlight the importance of interface quality for the SMR. Finally, using a spin diffusion model for the SMR response, we achieve reasonable agreement between simulation and experiment. Our results demonstrate how SMR experiments in trilayers can be used to study pure spin currents in materials with vanishing spin Hall angle $\theta_\mathrm{SH,NM}$.

\section*{Supplementary Material}
See supplementary online material for the structural and magnetic characterization of the bi- and trilayer samples, the carrier concentration of the Ga:ZnO layer, and a more elaborate discussion of the SMR trilayer simulation.

\begin{acknowledgments}
M.S.R. and J.M. would like to thank for funding from Department of Science and Technology, New
Delhi, that facilitated the establishment of Nano Functional Materials Technology Centre (Grant: SR NM/NAT/02-2005). J.M. would like to thank UGC for SRF fellowship. We acknowledge financial support by the German Academic Exchange Service (DAAD) via project no.~57085749.
\end{acknowledgments}
\bibliography{Biblio}

\beginsupplement
%

\section{Structural and magnetic characterization}

We first compare the results obtained for a Bi:YIG(54)/Pt(4) bilayer and a Bi:YIG(54)/Ga:ZnO(8)/Pt(4) trilayer, where the numbers in parentheses give the  film thicknesses in nm. The structural quality of the thin films was analyzed by x-ray diffraction (XRD). The magnetic properties were obtained by superconducting quantum interference device magnetometry.

\begin{figure}[t]
 \includegraphics[width=85mm]{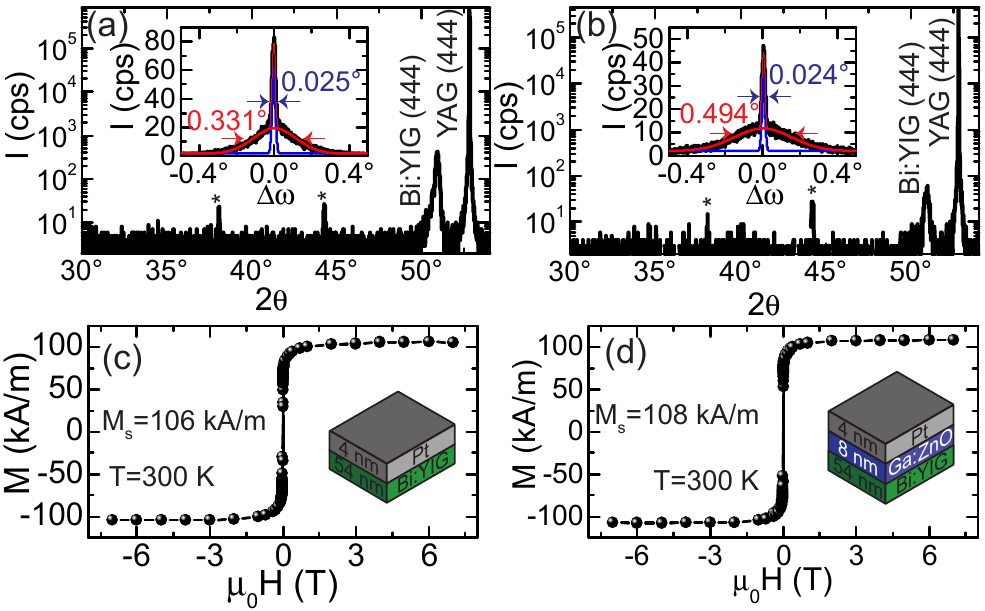}\\
 \caption[XRD and SQUID magnetometry results for Bilayer and Trilayer]{Structural and magnetic properties of the laser-MBE grown Bi:YIG(54)/Pt(4) bilayer and Bi:YIG(54)/Ga:ZnO(8)/Pt(4) trilayer samples grown on YAG (111) substrates. $2\theta-\omega$ scan of bilayer (a) and trilayer (b) samples, the insets of (a) and (b) are the rocking curve of the Bi:YIG (444) reflection, numbers represent the FWHM of the corresponding Gaussian fit.  Peaks marked with asterisks are background reflections from the sample holder. In-plane magnetization versus applied magnetic field curve of bilayer (c) and trilayer (d) samples measured at $300\;\mathrm{K}$.}
  \label{figure:XRD_BiYIG}
\end{figure}

The XRD results presented in Fig.~\ref{figure:XRD_BiYIG} for the bilayer (panel a) as well as for the trilayer (panel b) indicate excellent structural quality. In both samples we only observe reflections that can be assigned to the substrate or the layers itselfthemselves. In the trilayer sample no reflections of the Ga:ZnO layer could be observed, which we attribute to the small thickness of the layer. The XRD rocking curves of the Bi:YIG (444) reflection (insets in Fig.~\ref{figure:XRD_BiYIG}(a),(b)) consist in both samples of one rather broad peak superimposed on a second narrow peak. We attribute this to partial strain relaxation in the film due to the large lattice mismatch of $3\;\%$, between YAG and Bi:YIG. Nevertheless, the full width at half maximum (FWHM) extracted from Gaussian fits to the data yield nearly identical values of $0.025^\circ$ for the bilayer and $0.024^\circ$ for the trilayer sample. However, we obtain different values for the FWHM of the broader peak, $0.331^\circ$ for the bilayer and $0.494^\circ$ for the trilayer sample. The broad peak in the Bi:YIG (444) rocking curve can be attributed to the strain relaxation layer in the Bi:YIG forming at the YAG substrate interface. Taking this into account, we attribute the change in width to a possible change in strain relaxation by defect formation in the Bi:YIG layer, due to the different thermal treatment by the additional Ga:ZnO deposition in the trilayer. However, further investigations are necessary to confirm this assumption.

The magnetization curves recorded at $300\;\mathrm{K}$ are shown in Fig.~\ref{figure:XRD_BiYIG}(c) for the bilayer and (d) for the trilayer sample, where a diamagnetic background from the substrate has been subtracted from the raw data. We extract a saturation magnetization $M_\mathrm{s}=106\;\mathrm{kA/m}$ for the bilayer and $M_\mathrm{s}=108\;\mathrm{kA/m}$ for the trilayer sample. Both numbers are smaller than the bulk value~\cite{Hansen1983} $M_\mathrm{s}=144\;\mathrm{kA/m}$, which we attribute to defects in our layers. These defects can either be structural defects in the strain relaxation layer present at the YAG substrate interface~\cite{Popova2001}, effectively reducing the total saturation magnetization over the whole film thickness, or iron and oxide vacancies present in the whole Bi:YIG film~\cite{Dumont2005}. The coercive field for both samples is $0.6\;\mathrm{mT}$. Taken together, the structural and the magnetic properties of the bi- and trilayer sample are nearly identical. This suggests that the influence of the Ga:ZnO deposition on the magnetic properties of the Bi:YIG layer is negligible.

\section{Ordinary Hall effect in Ga:ZnO}

\begin{figure}[t]
 \includegraphics[width=85mm]{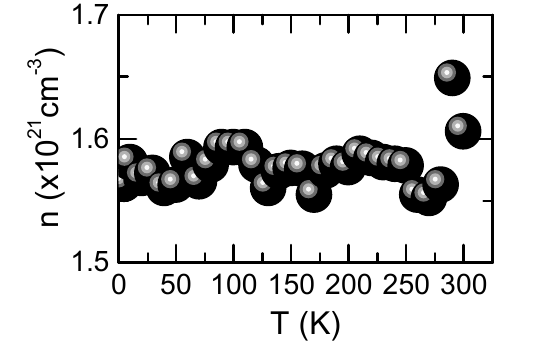}\\
 \caption[Ordinary Hall effect measurements]{Carrier concentration of a $8\;\mathrm{nm}$ thick Ga:ZnO grown on sapphire. The carrier concentration is independent of temperature indicating the degenerate doping of the Ga:ZnO layer.}
  \label{figure:carrierZnO}
\end{figure}

We used ordinary Hall effect (OHE) measurements on a $8\;\mathrm{nm}$ thick Ga:ZnO layer grown via pulsed laser deposition on a (0001)-oriented sapphire substrate under identical deposition conditions as for the trilayer structures investigated for the SMR experiments. From the OHE measurements, carried out at various temperatures, we then extracted the n-type carrier concentration $n$ as a function of temperature, the results are shown in Fig.~\ref{figure:carrierZnO}.

The carrier concentration is independent of temperature with $n\approx1.6\times10^{21}\mathrm{cm^{-3}}$. This indicates that the Ga:ZnO is degenerately doped.

\section{Simulation of the SMR in trilayer-systems}

\begin{figure}[t]
 \includegraphics[width=85mm]{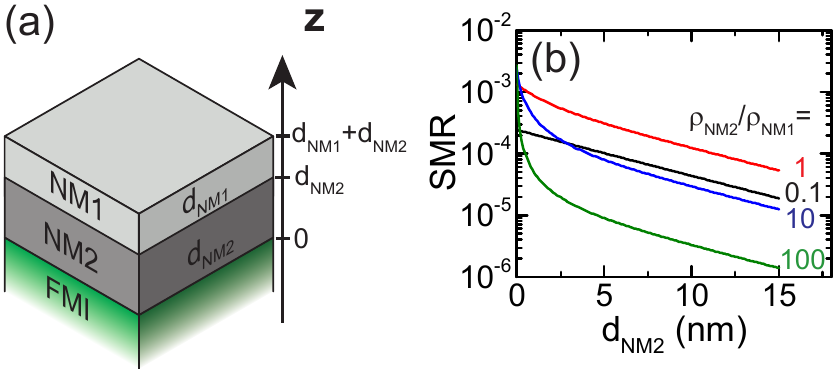}\\
 \caption[SMR model for trilayers]{Simulation of the SMR in a trilayer structure with two conductive layers and one FMI layer. (a) Schematic drawing of the coordinate system used for the calculations. (b) Dependence of the SMR on the thickness of the spin Hall inactive interlayer for $\rho_\mathrm{NM2}/\rho_\mathrm{NM1}=$ $0.1$ (black line), $1$ (red line), $10$ (blue line), and $100$ (green line). Similar to the conductivity mismatch problem for spin injection into semiconductors, the SMR effect is reduced.}
  \label{figure:SMRmodel}
\end{figure}

For modelling the SMR in trilayer structures we follow the approach outlined in Ref.~\onlinecite{chen_theory_2013}. For the calculations, we use the coordinate system depicted in Fig.~\ref{figure:SMRmodel}. Our system consists of two conductive layers (NM1, NM2), where the pure spin current in both layers is carried by the spin angular momentum of the charge carriers. In our calculations we assume that only the charge current $\mathbf{j}_{q\mathrm{,NM1}}(z)$ flowing in NM1 gets converted into a spin current $\mathbf{j}_{s,\mathrm{NM1}}(z)$, while there is no spin current generation via the spin Hall effect in NM2(spin Hall inactive layer). At the NM1/NM2 interface we assume that the spin-dependent electrochemical potentials $\bm\mu_{s\mathrm{,NM1}}(z)$, $\bm\mu_{s\mathrm{,NM2}}(z)$ and the spin current flowing across the interface are continuous, and thus neglecting any interface resistance contributions. This leads to the following boundary conditions:
\begin{eqnarray*}
           \bm{\mu}_{s,\mathrm{NM1}}(d_\mathrm{NM2}) &=& \bm{\mu}_{s,\mathrm{NM2}}(d_\mathrm{NM2}), \\
           \mathbf{j}_{s,\mathrm{NM1}}(d_\mathrm{NM2}) &=& \mathbf{j}_{s,\mathrm{NM2}}(d_\mathrm{NM2}), \\
           \mathbf{j}_{s,\mathrm{NM1}}(d_\mathrm{NM2}+d_\mathrm{NM1}) &=& 0, \\
           \mathbf{j}_{s,\mathrm{NM2}}(0) &=& 1/e \left[G_r \mathbf{m} \times (\mathbf{m} \times \bm\mu_{s,\mathrm{NM2}}(0)) + G_i ( \mathbf{m} \times \bm\mu_{s,\mathrm{NM2}}(0))\right],
\end{eqnarray*}
where $G_r$ and $G_i$ are the real and imaginary part of the spin mixing conductance per unit area.
We then solve the spin diffusion equation with the above boundary conditions as detailed in Ref.~\onlinecite{chen_theory_2013}. The analytical expressions obtained from this procedure are lengthy and therefore not written down in the text here. With these solutions, we then calculate the spin and charge currents as outlined in Ref.~\onlinecite{chen_theory_2013}. Averaging over the film thicknesses and expanding up to the second order in spin Hall angle we then obtain an expression for $\rho_\mathrm{long}$ of the whole trilayer stack. Using this expression we can then determine the SMR amplitude by determining $\rho_\mathrm{long}$ for $\mathbf{m}\parallel\mathbf{j}$ and $\mathbf{m}\parallel\mathbf{t}$. We note that similar calculations can also be carried out for a spin Hall active NM2 layer using the very same approach.

Using these result we can then calculate the NM2 thickness dependence of the SMR amplitude for different resisitivity ratios $\rho_\mathrm{NM2}/\rho_\mathrm{NM1}$, while using the following values for the remaining parameters: $g_{\uparrow\downarrow}=1\times10^{19}\;\mathrm{m^{-2}}$, $\lambda_\mathrm{sf,NM1}=1.5\;\mathrm{nm}$, $\theta_\mathrm{SH,NM1}=0.11$, $\lambda_\mathrm{sf,NM2}=12\;\mathrm{nm}$; The result of these calculations is shown in Fig.~\ref{figure:SMRmodel}(b). For $\rho_\mathrm{NM2}/\rho_\mathrm{NM1}=1$, the SMR gives the largest values for a finite thickness of NM2. For $\rho_\mathrm{NM2}/\rho_\mathrm{NM1}=0.1$, we find a lower SMR amplitude as compared to $\rho_\mathrm{NM2}/\rho_\mathrm{NM1}=1$ as now a large part of the electrical current runs through the spin Hall inactive layer and thus is not contributing to the SMR. For $\rho_\mathrm{NM2}/\rho_\mathrm{NM1}=10$ and $\rho_\mathrm{NM2}/\rho_\mathrm{NM1}=100$ the SMR gets enhanced for ultrathin NM2 layers, which can effectively understood as an enhancement of the interface resistance at the FMI/NM2 interface which boosts the SMR effect similar to the experiments on matching spin mixing conductance in spin pumping experiments as elaborated in Ref.~\onlinecite{Du2014}.

In our experiments we used Ga:ZnO as the spin Hall inactive NM2 to vary the spin diffusion length and resistivity of the NM2 layer as a function of temperature. Similar effects will occur if the carrier concentration is changed in the ZnO layer, as this will again affect the spin diffusion length and the resistivity of the NM2 layer.

\bibliography{Biblio}
\end{document}